# Geometrical Formulation of Quantum Mechanics

S. R. Vatsya

648 Inverness Ave., London, Ontario, Canada, N6H 5R4
e-mail: raj.vatsya@gmail.com  Phone: (1) 519 474 1183

## Abstract

Hamilton's action principle is formulated and extended in conformity with the gauge transformations underlying Weyl's geometry. The extended principle characterizes infinitely many equally likely trajectories with a particle traveling along a randomly selected one. Available similar formulations do not conform as directly to the gauge transformations as the present one. Also, they have not paid much attention to the path-independent, assigned, gauges. The freedom available in assigning these gauges is exploited here by defining them in terms of the configuration, and interactions of the observing system with the observed one. Impact of the method of observation on its outcome is described in terms of the assigned gauges so defined and illustrated with examples. A wavefunction is defined in a simply connected region essentially as an aggregate of the gauge transformations over all trajectories; equivalently, an aggregate of the Weyl-lengths acquired by a unit vector transported along all trajectories from everywhere. This representation is similar to Feynman's path integral representation differing only in that it incorporates the assigned gauges yielding an adjusted wavefunction that includes the impact of an observing system on the observed one. Probability density is shown to be a uniquely defined gauge invariant quantity but at the expense of the information about the observable effects contained in the gauge factors, assigned and otherwise. The particle trajectories defined here are thus shown to provide additional significant information about a system than provided by the wavefunction together with the probability density. Present description of the impact of method of observation on its outcome is compared with the descriptions according to the existing major interpretations of quantum mechanics.

## 1. Introduction

In the Riemann spaces, the length of a vector remains constant under parallel transport. The Weyl geometry [1] was developed by assuming that the length $l_x$ of a vector undergoes a change of $\delta l_x = a\phi'_\mu dx^\mu l_x$ under parallel transport from an arbitrary point $x$ to a neighboring point $x+dx$, where $a$ is a nonzero constant and $\phi'_\mu$, termed the Weyl gauge potentials, are components of a vector function $\phi'$ defined on trajectories in the underlying Riemannian manifold. It follows that the length $l_y$ at a point $y$ of a vector transported from $x$ to $y$ along a trajectory $\rho_{xy}$ is given by

$$l_y = \exp\left[a\int_{\rho_{xy}} \phi'_\mu\, dx^\mu\right] l_x. \tag{1}$$

Weyl argued further that the gauge $\kappa$ can be assigned arbitrarily at every point of the manifold, which is to recalibrate lengths by redefining the metric, $g(x) \to \kappa^2(x)g(x)$, transforming the length $l_x \to \kappa(x)l_x = l'_x$. It follows from (1) with this length recalibration,

$$l_y \to \kappa(y)l_y = l'_y = \kappa(y)\exp\left[a\int_{\rho_{xy}} \phi'_\mu dx^\mu\right]\kappa^{-1}(x)l'_x = \exp\left[\int_{\rho_{xy}} \left(a\phi'_\mu + \kappa^{-1}\kappa_{,\mu}\right)dx^\mu\right]l'_x, \tag{2}$$

where $,\mu$ denotes the derivative with respect to $x^\mu$.

A gauge transformation defined by (2), contains a path-dependent part, a functional, defined by (1), and a point function $\kappa(x)$. We refer to the path-dependent part as the essential gauge and to $\kappa(x)$, as the assigned gauge. This distinction is somewhat fuzzy as $\kappa(x)$, completely or in part, can be absorbed in $\phi'$, which cancels out in transports along the closed curves leaving the essential gauge but can alter the results where transport along unclosed curves is required. In view of this, $\phi'$ may have to be properly defined to be compatible with the physical system it describes, clearly distinguishing the essential and assigned gauges.

Weyl had originally introduced the concept of length recalibration and the parallel transportation rule $\delta l_x = \phi'_\mu dx^\mu l_x$ was introduced due to the resulting necessity. However, the two are essentially independent concepts as the metric and the affine connections in a general affinely connected space are independent. The present way of introducing the concept is clearer and more suitable for the present purpose, particularly as the quantum measurement will be described in terms of the assigned gauges. This is the reason for clearly distinguishing the essential and the assigned gauges from each other, which is usually not emphasized.

Attempting to deduce the basic formulation of quantum mechanics from the gauge transformations, London [2] took $a\phi'_\mu = -ie\phi_\mu$ and showed that with $y$ varying along the classical trajectory of a particle of charge $e$ in an electromagnetic field defined by the potentials $\phi_\mu$, $l_y$ is directly proportional to an extended de Broglie wavefunction associated with the particle. Adler [3] used the Weyl-London formulation to obtain Bohr's quantization condition. Wheeler [4] investigated the correspondence between Weyl's geometry and quantum mechanics in considerable detail. In particular, the amplitude associated with a physical system was expressed as a Wiener integral paralleling the Feynman path integral formulation of quantum mechanics [5-7]. With $\phi'_\mu = S_{,\mu}$, where $S$ is the classical action for a charged particle in an electromagnetic field in its nonrelativistic approximation, Wheeler deduced the Fokker-Planck diffusion equation. Nelson had earlier inferred the Schrödinger equation from this diffusion equation [8]. Thus, the two together link the gauge transformations and thus, Weyl's geometry, with quantum mechanics. Significant progress in this direction was made by a somewhat different approach, which was to express and extend Hamilton's action principle in conformity with the gauge transformations to deduce some quantum mechanical results [9,10].

Although significant, the above developments have only been partially successful in formulating the principles underlying quantum mechanics in terms of the gauge transformations. Furthering an earlier description [11], we describe here a formulation founded upon an extension of Hamilton's action principle that characterizes the particle trajectories. This formulation is directly compatible with the gauge transformations improving significantly upon the previous similar ones. The assigned gauges in Weyl's formulation are essentially arbitrary and in the related literature, they were considered largely inconsequential. Here this freedom is exploited by defining them in terms of the modifications to the underlying geometry resulting from an experimental arrangement, i.e., the observing system, interacting with the observed one, which together describe an act of observation. The extended action principle with physically assigned gauges is then used to describe the impact of an observation on its outcome.

A wavefunction in a simply connected region is defined from an adjusted form of Feynman's path-integral formulation, rendering it compatible with the extended action principle. This defines the wavefunction essentially as an aggregate of the gauge transformations over the particle trajectories yielding a wavefunction that micropores the assigned gauges and thus the impact of a mild observation. Strong intrusions of the observing system generate multiply connected regions. Wavefunction is naturally interpreted geometrically, as an aggregate of the Weyl lengths of a unit vector transported along all trajectories from everywhere.

Methods to estimate the values of assigned gauges and their observable effects are discussed in some experimental situations. Multiplicative essential gauge transformations arising in some phenomena are also known to produce physically observable effects [12,13]. The probability density is shown to be gauge invariant but at the expense of the additional information contained in the gauge or phase factors, assigned and the multiplicative essential ones.

The particle trajectories defined here augment the standard formulation of quantum mechanics improving upon the quantum description of the physical systems. In particular, the present formulation provides an improved understanding of quantum measurement, which differs radically from that in the usual interpretations of quantum mechanics.

## 2. Extended Action Principle

Hamilton's action principle assumes the action $S$ to be stationary about the particle path $\rho_{xy}$, i.e., $\rho_{xy}$ is an extremal defined by this variational characterization. The action principle is stated as

$$\delta S = \delta \int_{\rho_{xy}} dS = \int_{\rho_{xy}^{c,\text{inf}}} dS \cong 0,$$

where $\cong$ indicates that the equality holds up to the first order in area enclosed by each closed curve $\rho_{xy}^{c,\text{inf}}$ obtained as the union of an arbitrary curve $\rho_{xzy}$ in a small neighborhood of $\rho_{xy}$ and another arbitrary curve $\rho_{yz'x}$ in a small neighborhood of $\rho_{yx}$ with $x$ and $y$ fixed. Now the classical action principle can be expressed as

$$1 + a\,S\left(\rho_{xy}^{c,\text{inf}}\right) \cong \exp\left[a\,S\left(\rho_{xy}^{c,\text{inf}}\right)\right] \cong 1 \tag{3}$$

for all closed curves $\rho_{xy}^{c,\text{inf}}$ enclosing an infinitesimally small area with fixed $x$ and $y$ on the boundary, where $a$ is an arbitrary nonzero constant. From (2), The choice $\phi'_\mu = aS_{,\mu}$ defines the gauge group element associated with $\rho_{xy}^{c,\text{inf}}$ to be $\exp\left[a\,S\left(\rho_{xy}^{c,\text{inf}}\right)\right]$. Thus, (3) formulates the classical action principle in terms of the gauge transformations that constitute the foundation of Weyl's geometry.

Classical characterization of particle motion stated in (3) motivates its natural extension in the framework of the gauge transformations:

$$\kappa(y)\,\exp\left[a\,S\left(\rho_{xy}\right)\right]\,\kappa^{-1}(x) = 1, \tag{4}$$

which is a basic assumption underlying the present formulation. From (2), the left side of (4) is the gauge group element with $\phi'_\mu = S_{,\mu}$, associated with an arbitrary curve $\rho_{xy}$, closed or unclosed, and the equality is assumed to hold exactly. Thus, while (3) characterizes the classical trajectories in terms of the gauge group elements $\exp\left[a\,S\left(\rho_{xy}^{c,\text{inf}}\right)\right]$ associated with $\rho_{xy}^{c,\text{inf}}$, (4) is a characterization of the particle trajectories in terms of the group elements associated with an arbitrary curve $\rho_{xy}$. With the collection of curves $\{\rho_{xy}\}$ restricted to $\{\rho_{xy}^{c,\text{inf}}\}$, (4) reduces to (3) up to the first order, i.e., (3) is a special case of (4), and thus (4) is a proper extension of (3). Solutions of (4) facilitating the passage of a particle will be called the physical paths. Compared with earlier formulations [9,10], which are similar to but differ in some respects from the present one, the formulations of (3) and (4) conform directly to the gauge transformations.

Assigned gauges in the Weyl geometry are arbitrary. Since the assigned gauges for the closed curves cancel out, they have no effect on the descriptions based on such trajectories, e.g., the classical action principle and Bohr's orbits. However, if the description of a physical system involves other trajectories, e.g., Feynman's path integral formulation, the results can depend on the assigned gauges also.

We require that the assigned gauge $\kappa(x) = \kappa(x')$ if and only if the points $x$ and $x'$ in the underlying manifold are physically equivalent, constituting another basic assumption of the present formulation. For illustration, consider a free particle collimated by a small aperture and detected at some distance. The particle experiences the same physical conditions at all points in the interior of any path it may take and thus at all points except about the aperture and the detection point. However, it is subjected to different interactions at the aperture and at the detector, which are both different from the physical conditions everywhere else. Thus, the aperture and the detection points are physically inequivalent and also inequivalent to all the other points along the trajectory; therefore, the values of assigned gauges are different at the aperture, at the detector and at the other points of the relevant region where the gauge is constant. Present assumption about the assigned gauges thus, takes the experimental configuration and the involved interactions into consideration. This issue will be discussed further later.

It is clear from (4) that a constant value of the assigned gauges cancels out as is the case for a physical system confined to a region with all of its points being equivalent to each other, e.g., a free particle. In case of the closed curves, the assigned gauges are ineffective due to cancellations. In all such cases, (4) reduces to $\exp[aS(\rho_{xy})] = 1$. Since nonzero real values of the action for such particle trajectories cannot be excluded, $a$ should be purely imaginary, which can be seen by taking the logarithms of both sides of $\exp[aS(\rho_{xy})] = 1$. Thus, London's assumption of $a$ being purely imaginary [2], is deduced in the present formulation as a result from (4). In appropriate units, $a$ can be set equal to $i$. Since $a$ is the same constant for all cases, the characterization given by (4) reduces to

$$\kappa(y) \exp\left[i S\left(\rho_{xy}\right)\right] \kappa^{-1}(x) = 1. \tag{5}$$

Left side of (4) is the length acquired by a unit vector transported along $\rho_{xy}$. Thus, for a trajectory to be physical, i.e., an allowed path, (4) requires a vector attached to the particle to regain its length at some point along the path. Then it follows from (5) that the vector regains its length along such a path periodically. Adler [3] deduced Bohr's orbits from the Weyl-London construction by restricting the orbits to be circular and requiring the length to resume its original value as it returns to the original point. Adler's deduction can be seen to be a special case of (5). In the following, we show that the physically assigned gauges are expressible as phase factors reducing (5) further.

If a particle originating at $x$ is detected at $y$, then there must be at least one trajectory facilitating its passage. Let $\rho'_{xy}$ be such a trajectory. It follows from (5) that the assigned gauge at $y$ is given by

$$\kappa(y) \;=\; \kappa(x)\, \exp\!\left[-\,i\,S\!\left(\rho'_{xy}\right)\right],$$

implying that $|\kappa(y)|=|\kappa(x)|= const.$ for all relevant points $x$ and $y$. The argument is valid for each trajectory facilitating the passage of particle. Thus, $|\kappa(x')|$ can be taken to be a constant for all $x'$. It follows from (5) that the multiplicative constant in $\kappa$ cancels out, which can therefore be set equal to one. Since $|\kappa(x')|=1$ for all $x'$, $\kappa$ can be expressed as

$$\kappa(x') \;=\; \exp\!\left[-\,i\,\sigma(x')\right], \qquad (6)$$

with some function $\sigma(x')$, providing the phase factor representation of the assigned gauges.

From (6), the characterization given by (5) reduces to

$$\exp\!\left\{\, i\,\left[S(\rho_{xy}) \,-\, \sigma(y) \,+\, \sigma(x)\right]\right\} \;=\; 1, \qquad (7)$$

i.e.,

$$\left[S(\rho_{xy})-\sigma(y)+\sigma(x)\right] \;=\; 2n\pi$$

for all integers $n$. With the assigned gauge representation given by (6), (5) and (7) are equivalent characterizations of the physical trajectories, which will be referred to as convenient.

It is clear from (4), (5) and (7) that the assigned gauges are physically significant, i.e., they participate in describing the observable effects. This is one of the characteristics of the present formulation distinguishing it from the pertaining literature.

## 3. Physical Trajectories

While the action principle (3) defines a unique trajectory, its extension (5), equivalently (7), can be seen to have infinitely many solutions, i.e., there are infinitely many trajectories available for a particle to follow. Since there are no preferred paths, all are equally likely, and a particle trajectory would be determined by random selection. Thus, while a particle follows a definite trajectory, it is not determinable due to randomness. In the following, we describe some properties of the physical trajectories, which follow from (5) and (7).

If $\rho_{xy}$ and $\rho_{yz}$ are two physical paths, then the union $\rho_{xz}$ of $\rho_{xy}$ and $\rho_{yz}$ is also physical, which will be termed a continuing union. Physical trajectories can be classified under two categories: Monotonic and nonmonotonic. Along a monotonic trajectory, the action increases (decreases) monotonically as the path is traversed. If a trajectory is not monotonic, it is nonmonotonic. Nontrivial monotonic physical paths with smallest allowed value of the action in magnitude will be called the elemental physical paths. For a constant $\kappa$, the elemental paths $\rho_{xy}$ are defined by $S(\rho_{xy})=\pm 2\pi$. It is clear that all monotonic physical trajectories can be obtained by forming the continuing unions of the elemental paths. Thus, a particle travels along a randomly selected elemental from one point to the other, and then along another elemental from its arrival point to the next. It follows that a general monotonic particle trajectory is constituted of a sequence of randomly selected elementals. Nonmonotonic physical trajectories can be treated as the continuing unions of their monotonic segments, which by themselves may not be physical.

Consider a particle originating at a source and detected some distance away. The particle would travel along a continuing union of randomly selected elementals, which may or may not coincide with or be close to an extremal of the classical action principle. Thus, the particle would be detected as a localized entity, i.e., a particle, but not necessarily at the location determined by (3). If such an experiment is conducted with a large collection of identical particles or many times each with a single particle with particles in all repeats being identical, then each particle would travel along an independent continuing union of elementals. Consequently, the collection of the detected particles may or may not cluster about the location determined by (3).

Although the present formulation considers the entities to be purely particles, (7) imparts some wavelike propagation properties to their collective motion. Passage along randomly selected elementals can be seen to be similar to Huygens' construction for wave propagation, which can be visualized more clearly in a Euclidean plane with elementals restricted to the classical extremals. However, there is a significant difference: While the radii of the secondary wavelets in Huygens' construction are taken as convenient, in the present case the radii of the secondary surfaces that the elementals originating at a surface can reach are equal to the elementals. More significant and relevant property of the waves is the superposition principle, which we consider below.

If two identical waves originating at points $x$ and $x'$ meet at a point $\hat{x}$ in the same phase, then they interfere constructively with consequent amplitude being double of the amplitude of one of them; if they meet in opposite phases, then the resultant amplitude is equal to zero. The resultant amplitude varies in between the two extremes as the phase difference between them is varied.

In the present formulation, the left side of (5) is the length at $\hat{x}$ of a unit vector transported from its originating point, which from (7) is just a phase factor. For convenience, consider the case when $x$ and $x'$ are physically equivalent rendering the assigned gauges ineffective from (5). Then the unit vectors transported along $\rho_{x\hat{x}}$ and $\rho_{x'\hat{x}}$ acquire phases equal to $S(\rho_{x\hat{x}})$ and $S(\rho_{x'\hat{x}})$, respectively, with the phase difference being equal to

$$\Delta S = \left[ S(\rho_{x\hat{x}}) - S(\rho_{x'\hat{x}}) \right]$$

The pair of trajectories $\rho_{x\hat{x}}$ and $\rho_{x'\hat{x}}$ is equivalent to a nonmonotonic trajectory $\rho_{x\hat{x}x'}$. The action along $\rho_{x\hat{x}}$ increases and along $\rho_{\hat{x}x'}$, it decreases, yielding $S(\rho_{x\hat{x}x'}) = \Delta S$. If a unit vector transported along $\rho_{x\hat{x}}$ meets another unit vector transported along $\rho_{x'\hat{x}}$ in the same phase, then $S(\rho_{x\hat{x}x'}) = 2n\pi$ with $n$ being an integer. It follows in that case that $\rho_{x\hat{x}x'}$ is a solution of (5) implying that the particles can travel along $\rho_{x\hat{x}}$ and $\rho_{x'\hat{x}}$, which parallels, not duplicates, the constructive interference for the waves. If $S(\rho_{x\hat{x}'}) \neq 2n\pi$, the particles cannot travel along these trajectories. However, since $S$ is a continuous function on the trajectories, there are other trajectories in the neighborhoods of $\rho_{x\hat{x}}$ and $\rho_{x'\hat{x}}$ that can facilitate passage of the particles to a point in a small neighborhood of $\hat{x}$. In particular, if the two wavelets meet in opposite phases, the resultant intensity of two interfering waves is equal to zero. In the present formulation, if two unit vectors transported along a pair of trajectories meet in opposite phases at $\hat{x}$, then no particles would be transmitted by these trajectories but other paths can be found that can transmit particles to $\hat{x}$, resulting in a positive density of particles. Thus, there is a wavelike coherence built on the physical trajectories but only to a limited extent and the rules to calculate the particle density are not as straightforward as to calculate the intensity for the waves. However, this is sufficient for some arguments applicable to the waves to be applicable in the present case and for particles in a collection to exhibit some wavelike attributes as will be seen.

The superposition property of waves has been used to conclude that the waves of small wavelengths, equivalently with large phase lengths, propagate almost along the trajectories determined by requiring the phase length to be stationary [14, pp 34-36]. It is argued that as the trajectories are varied about an extremal, there is no change in the phase length up to the first order with respect to the variation, while the phase length changes in the first order as a path is varied about a nonextremal. Thus, a phase extremal has neighbors that meet at a point in the same phase or are slightly out of phase with consequent constructive, or almost constructive, interference and the neighbors of a nonextremal would be significantly out of phase resulting in a negligible intensity. This assertion is substantiated by calculations in some simple cases, which show rapid variation in the phase for longer phase lengths compared to shorter ones as a trajectory is varied about an extremal.

Essentially the same argument is used in Feynman's path integral formulation of quantum mechanics to conclude that the systems of a classical magnitude exhibit almost classical behavior, as follows [5, pp. 29-31]. In this formulation, an amplitude is associated with $x$ and $\hat{x}$ that is an equiweighted sum of the phase factors $\exp\{iS[\rho(\tau)]\}$ over all trajectories $\rho(\tau)$ from $x$ to $\hat{x}$. The phase factors associated with the trajectories about the extremals defined by a stationary $S$ are almost in phase yielding a higher contribution of these trajectories to the amplitude while the contribution from the trajectories away from the extremals adds up to a negligible amount due to large oscillations from trajectory to trajectory resulting from the first order changes in the action.

Although the above argument is applicable to show that in the present formulation, the macroscopic physical trajectories, i.e., the particle trajectories of a classical extent, are concentrated about the extremals, in the following we improve upon it for the present case.

Consider particles travelling along macroscopic trajectories. The action $S(\rho_{x\hat{x}})$ along a trajectory $\rho_{x\hat{x}}$ in an $\varepsilon$-neighborhood of an extremal $\rho_{x\hat{x}}^e$ is equal to $S(\rho_{x\hat{x}}^e)$ up to the first order in $\varepsilon$, i.e., $\left[S(\rho_{x\hat{x}}) - S(\rho_{x\hat{x}}^e)\right] = o(\varepsilon^2)$. Also, it follows from Hamilton's equations that $S(\rho_{(x+\delta)\hat{x}}^e) \simeq p_\mu \delta^\mu$ where $p_\mu = \partial L(\dot{x}, x)/\partial \dot{x}^\mu$ are the momentum components with $L(\dot{x}, x)$ being the Lagrangian defining the action. Since the action is a continuous function on the trajectories, there are pairs of trajectories $\rho_{x,\hat{x}}$ and $\rho_{(x+\delta)\hat{x}}^e$ such that $S(\rho_{x,\hat{x}}) = S(\rho_{(x+\delta)\hat{x}}^e)$ with some $\delta = o(\varepsilon^2)$. This argument is equally well applicable to the point $\hat{x}$. It follows that there are multiple pairs of equiaction trajectories in some $\varepsilon^2$-neighborhood of $\rho_{x\hat{x}}^e$ with at least one common end point. The action along nonmonotonic unions of such pairs is equal to zero, and thus, they are physical, since they satisfy (5). Therefore, the particles can travel along the monotonic segments of each such pair.

In many cases the action principle reduces to the minimum action principle. In such cases, the action increases continuously as a trajectory is varied about an extremal in any direction. Thus, there are pairs of equiaction trajectories flanking the extremal with their end points on it that can facilitate passage of the particles increasing the density. The same argument applies if the action principle is a maximum principle. Also, the neighborhood would be even smaller for large values of $p_\mu$. The neighborhood would also be smaller for macroscopic trajectories even for small values of $p_\mu$ as a slight deformation would result in a large variation in the action.

Existence of nonmonotonic physical trajectories can be shown about nonextremals also but spread over its $\varepsilon$-neighborhood. Also, some arguments favoring a higher concentration of the physical trajectories about an extremal if the action principle reduces to an optimum principle, do not apply to nonextremals.

It follows from the above that for a macroscopic system, that of a classical scale, the trajectories available to a particle would be concentrated about an extremal and thus it would travel along an almost classical trajectory. The argument applies, and the conclusion holds equally well for the monotonic segments of the nonmonotonic physical trajectories that are the continuing unions of their piecewise monotonic segments. As can be seen from (7), the assigned gauges have almost no impact on these conclusions.

Although the above argument is a significant improvement over the parallel arguments that have persisted in literature in cases of the waves and path integrals, it would still fall in the category of qualitative estimates and inferences. More accurate estimates and calculations are desirable in all cases, waves, path integrals and the present formulation. For the present case, quite accurate solutions of (5), (7) can be obtained numerically for the specific cases, which will be discussed further in Sec. 4. However, such calculations can only be case specific, and they would be rather long and somewhat tedious. Similar would be the case with improved estimates. Therefore, such calculations would be diversionary rendering them outside the scope of the present article where a founding background for further studies is presented. In general, the above arguments are sufficiently convincing to conclude that for a system of a classical scale, the physical trajectories should be concentrated about almost classical trajectories.

## 4. Role of Gauge in Observations

### 4.1. Double-Slit Experiment

Consider the double slit experiment where identically prepared particles, e.g., electrons, encounter two slits, at $x_i$ and $x_f$, in a screen and observed at an arbitrary location $\hat{x}$ on an observation screen [5, pp. 2-9]. If only one slit, e.g., at $x_i$ is open, all trajectories from $x_i$ to $\hat{x}$ are of a large extent and thus, the particles travel along macroscopic trajectories from the slit to the detector. As discussed in Sec. 3, these trajectories are concentrated about the extremals of the action principle implying that the particles travel along almost classical particle trajectories.

If both of the slits are open, then classically, the particles pass through the slits to reach about point $\hat{x}$ on the observation screen. Particle density about $\hat{x}$ is the sum of two densities each corresponding to its respective slit transmitted to $\hat{x}$ by the extremal joining the slit to $\hat{x}$, and no particles reach any other point. In the present formulation, a pair of trajectories meeting at $\hat{x}$ can be considered a continuing union of two monotonic trajectories $\rho_{x_i \hat{x}}$ and $\rho_{\hat{x} x_f}$ each of a classical extent constituting a nonmonotonic trajectory $\rho_{x_i \hat{x} x_f}$ from $x_i$ to $\hat{x}$ to $x_f$. In this arrangement, the assigned gauge at $\hat{x}$ cancels out reducing (5) to

$$\kappa(x_f) \; \exp\left\{i\left[S\left(\rho_{x_i \hat{x}}\right) - S\left(\rho_{x_f \hat{x}}\right)\right]\right\} \; \kappa^{-1}(x_i) \;=\; 1. \qquad (8)$$

If the points $x_i$ and $x_f$ are physically equivalent, which would be the case for the identical slits and particles, then $\kappa(x_i) = \kappa(x_f)$ and the solutions of (8) are given by

$$S \;=\; S(\rho_{x_i \hat{x} x_f}) \;=\; S(\rho_{x_i \hat{x}}) - S(\rho_{x_f \hat{x}}) \;=\; 2n\pi$$

with $n$ being an arbitrary integer. An extremal joining any two points in this case is a straight line with the momentum $p$ of the particle(s) being constant. Thus, if the monotonic segments $\rho_{x_i \hat{x}}$, $\rho_{\hat{x} x_f}$ of the corresponding nonmonotonic trajectory $\rho_{x_i \hat{x} x_f}$ are extremals of the action principle, then $S = p\Delta r'$, where $\Delta r'$ is the difference between the path lengths of the straight lines from $x_i$ to $\hat{x}$ and $x_f$ to $\hat{x}$. It follows that for a union of two straight lines to be physical, it is required that $\Delta r' = 2n\pi / p = \Delta r_n$. Since the physical trajectories are concentrated about the extremals $\rho_{x_i \hat{x}}$, $\rho_{\hat{x} x_f}$, a high concentration of the physical trajectories and hence of the particles should exist in the neighborhoods of such points, denoted by $\hat{x}_n$, decreasing away from them. In between the maxima, particles can reach only along nonextremals with small associated density. Incidentally, $\rho_{x_i \hat{x} x_f}$ is effectively a microscopic trajectory with the associated action equal to $2n\pi$.

The particle density distribution deduced above resembles the intensity distribution resulting from the interference of two identical wavelets emanating from $x_i$ and $x_f$. Essentially the observations of this type constituted the basis of the doctrine of wave-particle duality in quantum mechanics. The present alternative explanation is based on the characterization of the particle trajectories stated in (5).

Converse is also true, i.e., if $\hat{x}_n$ are accessible to the unions of piecewise extremals, then from (8), $\kappa(x_f) = \kappa(x_i)$. If the points $x_i$ and $x_f$ are not physically equivalent, then the assigned gauges $\kappa(x_i)$ and $\kappa(x_f)$ are not equal to each other. In this case, the particles can only be carried to these points by the physical trajectories that are not piecewise extremals reducing the density about them as discussed in Sec. 3. From (6) and (7), the degree of reduction would depend on the difference between the phases $\sigma(x_i)$ and $\sigma(x_f)$ of $\kappa(x_i)$ and $\kappa(x_f)$, respectively. Although $\hat{x}_n$ are not accessible to the corresponding piecewise extremals in this case, some piecewise extremals can transmit particles to some other points shifting the pattern.

One way to destroy the equivalence of $x_i$ and $x_f$ is by an intrusive observation behind the slits, e.g., by scattering photons from the particle beam. Again, we consider the extremals and the corresponding estimates, which are sufficient to glean qualitative information about the impact of such a measurement on the behavior of particles. Consider a beam of photons of momentum $p^{ph}$ parallel to the plane of slits scattered from the particles of momentum $p$ after they have passed through one of the slits about a point close to $x_i$. Scattering would transfer the momentum $\delta p = o(p^{ph})$ to the particles about $x_i$, which would shift the location of the corresponding beam on the observation screen. It can be seen that the momentum of the shifted beam in the direction of its propagation changes by about $(\delta p)^2/(2p)$. Since the distance $d_s$ between the slits is small compared to the distance $d_o$ between the planes of the slits, and the observation screen, pathlengths of the particle beams can be assumed to be almost equal to $d_o$. Thus, the action of the corresponding beam is altered by

$$\delta S \approx \left[ d_o (\delta p)^2 / (2p) \right] \approx s \delta p / 2,$$

where $\approx$ denotes "about equal to," and $s \approx d_o \delta p / p$ is the shift of this particle beam, i.e., its separation from the other beam on the observation screen. This is equivalent to altering the gauge $\kappa(x_i)$ to $\kappa'(x_i) \approx \kappa(x_i) \exp(i \delta S)$.

For small values of $s$, the particle beams can still be considered almost correlated since there would still be many pairs of almost extremals with a common point on the observation screen in the neighborhoods of $\hat{x}_n$. From (7) and the above discussion, this implies the existence of a shifted and smeared density pattern. As the scattering effects increase, the shift increases shifting and smearing the density distribution further. It is clear that when these effects become significant enough so that the resulting shift $s \geq d/2$, where $d = (\hat{x}_{n+1} - \hat{x}_n)$, i.e., the distance between the two consecutive maxima, the correlation between the beams is almost completely lost. Consequently, $\kappa(\hat{x})$ cancels out for almost no trajectories invalidating (8). Instead, the beams travel independently, which are shown in Sec. 3 to be almost classical particle trajectories yielding the associated classical observation. This associates the existence of an interference-like density distribution and the degree of correlation of the beams with each other, which is determined by the closeness of the phases of $\kappa(x_f)$ and $\kappa'(x_i) \approx \kappa(x_i) \exp(i\delta S)$. It is clear that this discussion is valid as long as $s$ does not exceed $(d/2)$ by a significant amount, which places a limit on $|\delta S|$ and $\kappa'(x_i)$ also.

It follows from purely geometrical considerations that $(d_s / d_o) = (\Delta r / d)$, where $\Delta r = (\Delta r_{n+1} - \Delta r_n)$ for each $n$. Since the transferred momentum is small compared to $p$, (8) still yields $\Delta r \approx 2\pi / p$. In the standard quantum mechanics, the corresponding results are $(d_s / d_o) = (\lambda / d)$ and $\lambda \approx 2\pi / p$, the wavelength of wave associated with the particle. Exact equalities would hold in both cases if there is no transfer of momentum. In both cases, we have $(d_s / d_o) \approx (2\pi / pd)$, implying that $s \approx (d_s \delta p / \pi)(d/2)$. Thus, for the existence of an interference-like particle density distribution, one should have $(d_s \delta p / \pi) < 1$. We take a statistical average $\delta p \approx p^{ph}/2$ for the amount of transferred momentum, which is compatible with the estimates being used here. The resulting condition for the existence of an interference-like density distribution reduces to $d_s < 2\pi / p^{ph}$.

The pathlength of an elemental extremal for a photon of momentum $p^{ph}$ is equal to $2\pi / p^{ph}$, which is the precision limit of the accuracy of its location. In the standard quantum mechanics, $2\pi / p^{ph}$ is equal to the wavelength of photon, placing the same limit on the precision of its location. If this pathlength is greater than $d_s$, the distance between the slits, then the slit through which a particular particle passed through cannot be determined. Thus, the loss of "which path" information is equivalent to $d_s < 2\pi / p^{ph}$, which is shown above to imply the existence of an interference-like density distribution. If $d_s > 2\pi / p^{ph}$, "which path" information becomes available and since then $s \geq (d/2)$, an interference-like density pattern does not exist; the beams are no longer correlated and thus, they travel almost as independent classical beams producing the classically expected density distribution for particles on the observation screen.

Similar arguments and estimates are used for about the same deductions in standard quantum mechanics [5, pp. 9-14] but with $\lambda \approx 2\pi / p$, a wave characteristic, which is eliminated in the present treatment. Distribution of the assigned gauges is determined by the geometrical configuration and their values depend on the pertaining interactions. Since the underlying manifold and the geometrical configuration in this case are quite precise, these estimates of the assigned gauges can be improved upon by more careful calculations, e.g., quite accurate solutions of (5) and (8) with net gauge effect being null can be obtained numerically for particles, including photons, of known momenta. Alteration to the physically assigned gauges for the precise systems can also be determined more accurately by the classical calculations, e.g., with the electrons and photons of given momenta, which should capture the bulk of its effect. For the reasons mentioned in Sec.3, such calculations are outside the scope of the present paper. Also, theoretical and experimental studies for similar systems so far have focused mainly on the existence and nonexistence of the interference-like density distribution. As discussed above, in the present formulation there should be a gradual transformation from one distribution pattern to the other, which should be discernible by the commensurate experimental studies providing a comparison between the calculated and the observed behavior.

The arguments used to illustrate an impact of the assigned gauge in the simple system considered above, remain valid in more complicated ones. In any experimental set up, there are some points, which are not equivalent to each other and with other points in the underlying manifold, e.g., at the reflectors, beam-splitters and the like, usually placed along the paths of the particles in the experiments designed to observe their behavior [15]. It can be inferred from the above discussion that a set of trajectories transmitting the particles to an observation region would produce an interference-like density distribution there if and only if the net gauge effect is insufficient to enable an extraction of "which path" information. If the "which path" information is available at the terminal point of a trajectory, it cannot be common to the corresponding pairs of the beams. As discussed above, then the beams would travel almost as independent classical beams. More detailed and accurate numerical calculations can be carried out on all such experiments in the same way as for the simple double slit experiment. It is pertinent to remark that the particle behavior here is described in terms of the correlation of trajectories, which is a geometrical concept, in contrast with the entangled physical systems, or particles, elsewhere [15]. These descriptions are interchangeable in practical terms, but they differ in their conceptual contents.

### 4.2. Aharonov-Bohm Effect

In classical mechanics, the electromagnetic potentials were introduced as mathematical auxiliaries to construct the fields that were still considered to describe the electromagnetic phenomena completely. However, the solutions of quantum mechanical equations depend directly on the potentials, which can produce measurable effects as was noticed by Aharonov and Bohm for an electromagnetic phenomenon [12].

To describe the Aharonov-Bohm effect, consider an experimental set up in which a controlled magnetic field is generated by a long vertical solenoid by controlling the current through its coil. The experiment is conducted close to one of its ends mimicking a magnetic monopole. An electron beam is split in two at a point $x_i$. One of the beams travels along a trajectory $\rho_1$ on one side of the cylinder and the other, along $\rho_2$ on the opposite side, both shielded from the magnetic field. Then the beams meet at another point $\hat{x}$. Thus, the union of $\rho_1$ and $\rho_2$ encloses the magnetic field but neither of the beams passes through it. The wavefunction representing an electron in this set up is given by $\psi = \psi_0 \exp(-ie\mathcal{S}[\rho])$, where $\psi_0$ is the wavefunction of a free electron and

$$\mathcal{S}(\rho) = \int_\rho \phi_\mu \, dx^\mu ,$$

where $\rho$ is a trajectory from $x_i$ to $x$ [12,13]. In a simply connected region, $\psi$ is uniquely defined, but in a multiply connected region as the region outside the cylinder, $\psi$ is multiple valued unless $\mathcal{S}(\rho)$ is an integral multiple of $2\pi/e$. In general, the wavefunction can be expressed as a linear combination of its two independent solutions:

$$\psi_1 = \psi_0 \exp[-i\,e\,\mathcal{S}(\rho)] \text{ and } \psi_2 = \psi_0 \exp[-i\,e\,\mathcal{S}(\rho')],$$

which represent the beams on the opposite sides of cylinder each staying in a simply connected region. Each of the solutions is valid in the respective simply connected region. It follows that the phase difference between the two beams meeting at $\hat{x}$, i.e., between $\psi_1$ and $\psi_2$ at $\hat{x}$, is equal to

$$e\left[\mathcal{S}(\rho_1) - \mathcal{S}(\rho_2)\right] = e \int_{\hat{\rho}} \phi_\mu \, dx^\mu = ef, \tag{9}$$

where the integral is taken along the closed curve $\hat{\rho}$ traversed from $x_i$ to $\hat{x}$ along $\rho_1$ and from $\hat{x}$ to $x_i$ along $\rho_2$, yielding the enclosed magnetic flux $f$ [13].

The phase difference being directly proportional to $f$, can be controlled by varying the current in solenoid. It is clear that if $f$ is an integral multiple of $2\pi/e$, then the two beams interfere constructively and if $f$ is an odd integral multiple of $\pi/e$, the two beams interfere destructively. It follows that the interference pattern produced by the two waves on the observation screen shifts as $f$ is varied, repeating itself periodically with the period $2\pi/e$ with respect to $f$.

It is clear that the solutions arising here are related by the appropriate multiplicative gauge transformations, which were termed the non-integrable phase factors [13]. Since the field was still considered directly responsible for the electromagnetic effects, early experimental observations of this effect [16] were looked upon with suspicion. However, in view of the subsequent observations [17], the effect is now well established experimentally. The phase factors were thus seen to describe electromagnetism adequately and completely instead of the fields or the potentials.

Having sketched above the standard argument to describe the Aharonov-Bohm effect in quantum mechanics, we describe it according to the present formulation. Consider Chambers' experiment [16], where one collimated beam is reflected at a point $x_1$ on one side of the cylinder and the other one is reflected at $x_2$ on the other side, both beams originating at $x_i$ and meeting at $\hat{x}$. The particle density about $\hat{x}$ can be estimated sufficiently accurately by considering the piecewise extremals, since they are macroscopic in extent. Since $\hat{\rho}$ is a continuing union of the trajectories from $x_i$ to $x_1$, $x_1$ to $\hat{x}$, $\hat{x}$ to $x_2$ and $x_2$ to $x_i$, the gauges assigned by the experimental set up at $x_i$, $x_1$, $x_2$ and $\hat{x}$ are rendered ineffective. This arrangement is treated the same way as the double slit arrangement with no net gauge effect.

Classically, the particles can be described by the Lagrangian

$$L = \left(-m \sqrt{\dot{x}_\mu \dot{x}^\mu} + e \phi_\mu \dot{x}^\mu\right).$$

As in case of the double slit arrangement where the physical trajectories are given by (8), the physical trajectories for the present arrangement are given by $S = 2n\pi$, where

$$S = S(\rho_1) - S(\rho_2) = \int_{\hat{\rho}} d\tau \left[-m \sqrt{\dot{x}_\mu \dot{x}^\mu} + e \phi_\mu \dot{x}^\mu\right] \qquad (10)$$
$$= S_0 + e \int_{\hat{\rho}} \phi_\mu dx^\mu = S_0 + e f.$$

Consider the case with $f = 0$. The physical trajectories for this case are defined by $S_0 = 2n\pi$. Let $\rho_1$ be the continuing union of straight lines from $x_i$ to $x_1$ and from $x_1$ to $\hat{x}$. Similarly, let $\rho_2$ be the continuing union of straight lines from $x_1$ to $x_2$ and from $x_2$ to $\hat{x}$. Since $\rho_1$ and $\rho_2$ are the continuing unions of extremals, physical particle trajectories are concentrated about the consequent physical trajectory $\hat{\rho}$. The momentum along each straight-line segment in the direction of propagation can be taken to be the same constant $p$. Free particle part of the action is given by $S_0 = p\Delta r'$, where $\Delta r'$ is the difference between the path lengths of $\rho_1$ and $\rho_2$. Therefore, the condition for $\hat{\rho}$ to be physical and hence, for the existence of a maximum is again $\Delta r' = 2n\pi / p = \Delta r_n$, yielding a particle density distribution resembling an intensity pattern produced by two identical interfering waves. As $f$ is varied, it follows from $S = S_0 + ef = 2n\pi$ that the pattern shifts repeating itself with period $2\pi/e$ with respect to $f$. In the present formulation, the Aharonov-Bohm effect is thus described solely by the present definition of the physical trajectories, in a sharp contrast with the standard argument based on the interfering waves.

The double-slit experiment and the Aharonov-Bohm effect share an essential similarity. In both of the cases, the observed effect depends on the fact that the two beams are related with each other through a multiplicative gauge or phase factor resulting from an alteration to the underlying geometry. In the former, this alteration results from the two slits generating an assigned gauge factor and in the later, from a hole punctured by the field confined almost to a point, generating an essential gauge factor.

## 5. Path Integral Formulation

Wheeler [4] pointed out that no unique trajectory can be assigned to a particle in Weyl's geometry, since a trajectory can be altered by altering the gauge. Then assuming that there are infinitely many paths available to a particle, a Wiener integral representation of an aggregate of the gauge transformations over all trajectories was developed. This parallels Feynman's path-integral formulation of quantum mechanics with the aggregate being the amplitude or wavefunction. Taking $\phi'_\mu = S_{,\mu}$ with Weyl's constant $a$ being real, together with a set of assumptions, Wheeler deduced the Fokker-Planck diffusion equation for a nonrelativistic charged particle in an electromagnetic field, which was known to be related to the corresponding Schrödinger equation [8]. This associates Weyl's geometry with quantum mechanics although to a limited extent.

In addition to having a limited scope and encumbrance of numerous assumptions, Wheeler's formulation lacks accessibility and it does not provide much understanding of the particle motion. Also, in its present form, it has not proven to be very suitable for deducing the quantum mechanical equations. As discussed above, the present formulation of the particle trajectories by (5) yields a clearer picture of the particle motion and several quantum mechanical results in a straightforward manner. This also enables a path integral representation of the wavefunction in close similarity with Feynman's, as done below.

For a unit vector, i.e., $l_x = 1$, the left side of (2) with $a\phi'_\mu = iS_{,\mu}$ and constant gauge, is a typical term $\exp\{i\, S[\rho(\tau)]\}$ in Feynman's representation of the amplitude [5, pp. 28-29]. In view of this observation, Feynman's quantum mechanical amplitude for a particle to go from $[x_0, \tau(x_0)]$ to $[x, \tau(x)]$ is naturally adjusted to read

$$\mathcal{K}\{\kappa; [x, \tau(x)], [x_0, \tau(x_0)]\} = \sum_{\substack{\text{all paths} \\ \text{from } x_0 \text{ to } x}} \kappa(x)\, \exp\{i\, S[\rho(\tau)]\}\, \kappa^{-1}(x_0)\,, \qquad (11)$$

where $S[\rho(\tau)]$ is the classical action along the trajectory $\rho(\tau)$ in the underlying manifold joining the two points parameterized by a suitable parameter $\tau$. Terms in (11) are obtained by replacing the Feynman terms $\exp\{i\, S[\rho(\tau)]\}$ by the left side of (5). The sum in (11) assigns equal weight to each term, which is absorbed in it.

Feynman had obtained the formulation for a nonrelativistic system in 3D Euclidean space with time being the parameter and deduced the Schrödinger equation; alternative forms, e.g., for the propagator instead of the amplitude, and generalizations were developed later [6,7]. The generalized form (11) is valid on the background of a Riemannian space where the parameter is taken to be the arclength treated as an independent parameter [6,7,18,19, arXiv1405.7693v.1]. The formulation (11) is expressed for the amplitude for its suitability for the present purpose.

Due to the periodicity defined by (5), the sum in (11) reduces to that over the elemental physical paths containing the point $x$. The wavefunction is defined by (11) by letting $x_0$ arbitrary [5, pp. 57-58]. For a gauge free manifold, i.e., with $\kappa$ being a constant, which then cancels out, (11) reduces to the Feynman sum. Thus, (11) is just a slightly adjusted form of Feynman's formulation but with a significant difference in that it includes the assigned gauges that incorporate the impact of an observation on the system.

The representation of (11) assumes that the trajectories $\rho(\tau)$ are gauge neutral in their interiors, i.e., $\kappa$ can be taken to be equal to one there, which includes the cases of nonconstant $\kappa$ as long as each $\rho(\tau)$ remains a continuing union of its segments. This holds true in cases where an intrusion altering the assigned gauge distribution is not strong enough to decohere the system of trajectories significantly. As seen in Sec. 4.1, some measurements involving strong intrusion can break the continuity disjointing the sets of trajectories. In such cases, the representation of (11) remains valid in each region where the stated condition is satisfied and (11) would yield a different wavefunction in each such region. Thus, as is the case with Feynman's original representation, present one remains valid in each simply connected region. If the underlying manifold itself is not simply connected but still is a union of more than one such regions as in Sec. 4.2, then (11) yields a multiple valued wavefunction. Since the Klein-Gordon equation is deducible by the path integral method [18,19, arXiv1405.7693v.1], its solutions considered in Sec 4.2 are obtained essentially from the Feynman form of (11). The results of the present section based on (11) are valid for mild intrusions and simply connected regions with similar results for each such region.

It follows from (2) that with $a\phi'_\mu = iS_{,\mu}$, each term in (11) is equal to the length at $[x, \tau(x)]$ of a unit vector transported from $[x_0, \tau(x_0)]$ along $\rho(\tau)$. Since the wavefunction is defined by (11) with $x_0$ being arbitrary, it defines the wavefunction as an aggregate of the Weyl lengths at $[x, \tau(x)]$ of a unit vector transported to this point along all physical trajectories from everywhere; where the Weyl length is defined by (2) with $a\phi'_\mu = iS_{,\mu}$. This provides a geometrical interpretation of the wavefunction.

As discussed before, a particle follows a definite trajectory out of many but because of randomness, it is undeterminable. An aggregate of the Weyl lengths over all particle trajectories defined by (11) provides a description of the particle's behavior complementing (5). Even though the wavefunction may represent a single particle, in view of its definition by (11), it incorporates the properties of infinitely many particles; equivalently, identical particles one in each of infinitely many equivalent settings; and it does not contain a detailed information of the motion.

Although Wheeler's and Feynman's formulations are based essentially on the particle picture, as is the present one, there are significant differences, as follows. Neither of the former two defines any particle trajectories and each one assigns all trajectories to a particle by assumption. Wheeler also assumes that the probability of a particle following a trajectory is inversely proportional to the change in length produced by an infinitesimal displacement along the trajectory and thus, inherently assigns different probabilities to the trajectories. Concept of the gauge transformations does not enter Feynman's formulation and it assumes that a particle takes all alternative paths each with equal probability. In comparison, the present formulation is based on a purely particle picture assigning infinitely many equally likely particle trajectories characterized by (5), which yields correct description of the behavior of particles. Wheeler`s and Feynman`s formulations do not consider the issue of impact of the method of observation on its outcome, which is described successfully in the present formulation, in Sec. 4.1. Weyl's weights used by Wheeler are not assumed to have any physical meaning. Also, Wheeler's formulation does not provide a clear interpretation of the wavefunction in Weyl`s geometry. In comparison, the gauge transformations in the present formulation are defined in terms of the assigned gauges with clear physical significance as they are determined by the physical properties of the observing system interacting with the observed one. Both definitions of the gauge transformations are mathematically equivalent, but the present one has advantages as discussed above. Also, the present formulation provides a clear geometrical interpretation of the wavefunction, in terms of the Weyl lengths.

Following the standard procedure of path integration to express the sums in integral forms [5-7], (11) is expressed as

$$\kappa^{-1}(x)\mathcal{K}'(\kappa;x,\tau+\varepsilon) = \int d\hat{m}(y)\exp\left[i\,\hat{S}(x,\tau+\varepsilon;y,\tau)\right]\kappa^{-1}(y)\mathcal{K}'(\kappa;y,\tau); \quad (12)$$

where $\mathcal{K}'(\kappa;x',\tau') = \mathcal{K}\{\kappa;[x',\tau'],[x_0,\tau(x_0)]\}$; $\hat{m}(y)$ is a suitable measure, which depends in part on the geometry of the underlying manifold; $\tau$ is the arclength treated as an independent parameter, $\hat{S}(x,\tau+\varepsilon;y,\tau)$ is the action along the extremal joining $[y,\tau]$ to $[x,\tau+\varepsilon]$ as $y$ varies over the manifold with fixed $\tau = \tau(y)$, and $(\tau+\varepsilon) = \tau(x)$. The wavefunction $\psi$ is defined by letting $x_0$ vary over all points in the underlying space, yielding

$$\kappa^{-1}(x)\psi(\kappa;x,\tau+\varepsilon) = \int d\hat{m}(y)\ \exp\left[i\,\hat{S}(x,\tau+\varepsilon;y,\tau)\right]\kappa^{-1}(y)\ \psi(\kappa;y,\tau). \quad (13)$$

For a constant gauge, (13) reduces to Feynman's representation of the quantum mechanical wavefunction $\psi$:

$$\psi(x,\tau+\varepsilon) = \int dm(y)\ \exp\left[i\,\hat{S}(x,\tau+\varepsilon;y,\tau)\right]\psi(y,\tau), \quad (14)$$

which is the same as (13) with $\psi$ replacing $\kappa^{-1}\psi(\kappa)$ implying that $\kappa^{-1}\psi(\kappa) = \psi$. Thus, the wavefunction $\psi(\kappa)$ representing the basic physical system together with the impact of the observing system can be constructed by multiplying $\psi$ by the applicable assigned gauge. It is clear that the wavefunction and the characterization of the physical trajectories given by (5) together describe the motion more completely than either one alone.

In the present formulation, an act of observation assigns certain gauges to points in the underlying manifold. Thus, a measurement alters $\psi$ to $\psi(\kappa)$, which depends on the observing system and the interactions involved. Clearly, the wavefunction can be altered by altering the details of an observation. Therefore, the concept of an unambiguous measurement cannot be assigned to a wavefunction. To eliminate the ambiguity, an associated gauge invariant quantity should be constructed and measured for a physical measurement to be objective and meaningful. Wheeler constructed the conjugates of tensors in Weyl's geometry with the product of the tensor and its conjugate being a quantity of vanishing Weyl weight, which was assumed to be physically measurable. In the present formulation, it follows from (6) that $\kappa^{-1}(x) = \kappa^*(x)$, the complex conjugate of $\kappa(x)$, implying that $\psi^*\psi = (\kappa\psi)^*(\kappa\psi)$ is a real gauge invariant quantity, rendering $\psi^*$ a natural conjugate of $\psi$. Born's probability density $\psi^*\psi = |\psi|^2$ is clearly a uniquely defined gauge invariant quantity with a clear outcome of any measurement regardless of the gauge but at the expense of the information contained in the gauge factors, both assigned and the multiplicative essential, as discussed in Sec. 4.

## 6. Quantum Measurement

There are number of interpretations of quantum mechanics, each with its own explanation of the impact of an observation, or the observer, on what is observed. Neither one of them has gained universal acceptance; each one has its supporters and critics with varied views, which have changed with time. The interpretations together with their critiques are described in detail in literature. Here we discuss a few major ones in brief to compare the understanding of the indicated phenomenon according to the present formulation with parallel views in the interpretations of quantum mechanics. Considerations are restricted to the double slit experiment discussed in Sec. 4.1.

Copenhagen interpretation holds that the system is represented by a wavefunction, which collapses in one particular state consequent to a measurement, in the state that is observed. This view assigns dual nature, both wave and particle, to each of the entities, having no definite form until observed. Thus, the method of observation in a sense "bestows" form upon the entity observed. In comparison, the present formulation assumes a definite form, particle, for each of these entities and each one is expected to be observed as a particle individually. In a bulk, the entities' behavior in the gauge neutral case resembles that of a wave in the sense that the particle density distribution resembles the intensity distribution of two interfering wavelets. As discussed in Sec. 4.1, if the intrusion by the observing system is sufficiently strong to yield a definite outcome, it destroys the correlation between trajectories from the two slits and the set of trajectories breaks up in two; one of them corresponding to the observation region and the other, covering its complementary region. The deduced observation then compares with a collection of classical particles in agreement with the observed behavior. As discussed in Sec. 5, separate wavefunctions can be constructed for each of the regions. Thus, the wavefunction $\psi$ representing the particles from both slits in the gauge neutral case "breaks up" into two fragments. The fragment corresponding to the observation region can be considered the "collapsed" form of $\psi$ providing an interpretation of the "wavefunction collapse."

The Bohm formulation [20] or the pilot wave interpretation proposed by de Broglie, assumes that these entities are particles and each particle follows a definite trajectory but guided by a probability wave determined by the state of the system. Thus, a particle passes through one of the slits, but the wave passes through both, and the probability of a particle being at a location on the observation screen is proportional to the resultant intensity of the interfering pilot wavelets at that location. Thus, the particle density pattern resembles the intensity distribution of two interfering wavelets without the particle assuming a wave form. Particle trajectories remain unknown, which are considered the hidden variables. Concerning the collapse of a wavefunction, the Bohm theory considers a universal wavefunction incorporating the observed and observing systems together as a composite system, which never collapses; collapse occurs only in the phenomenological sense applicable to a subsystem caused by decoherence resulting from an intrusion by a measurement; and the subsystem evolves separately.

According to the present formulation, each one of the entities, assumed to be a particle, passes through one of the slits, which remains unknown but due to randomness. Also, while a definite trajectory cannot be assigned to a particle, sets of the physical trajectories can be computed from the information about the observing and the observed systems, and thus they are not completely hidden. The trajectories possess some wavelike properties to a limited extent but not the wave nature. Probability of finding a particle at a location is proportional to the density of paths available for passage of a particle. Present explanation of the observation in the double slit experiment is discussed in Sec. 4.1; and interpretation of "wavefunction collapse" is discussed above, differ radically from the Bohm's. Wavefunction of the composite system, the observed and the observing one together, does not enter the considerations.

Many worlds view [21] assigns purely wave nature to these entities. It was developed by coupling the observing system with the observed one through wavefunction of the composite system. Each measurement leaves each subsystem in an undetermined relative state with respect to the other. With each observation, system's state branches off to different non-communicating worlds, each one corresponding to each of the possible outcomes of an observation. What is observed depends on the world the observer enters. Such branching occurs at each subsequent observation. Thus, the outcome of a measurement depends on all previous measurements.

In the present formulation, the method of observation couples the two systems through the assigned gauges, which are at least approximately computable. Basic physical system and the assigned gauges determine the observation. Each act of observation assigns a multiplicative gauge factor altering the "state" of the observed system, which is altered further by each subsequent measurement. If an observation causes strong intrusion, the system of trajectories changes in a drastic way. As discussed above, one "branch" of the trajectories corresponds to the observation in the lab and the other, to its complement but in the same "world," and what is observed depends on which branch one observes. Thus, different parts of the same data can yield different outcomes, as observed [15]. Each one of these branches has an associated wavefunction. Such "branching" occurs with each measurement yielding a definite outcome. This provides the present interpretation of the "many worlds."

The concept of a wave is basic to the interpretations of quantum mechanics, which is irrelevant to the present formulation. Concept of the gauge transformations, which is fundamental to the present formulation, does not enter the interpretations.

In the present formulation, expected observation in the experiments involving large number of particles can be determined, to some extent, from the information about the observed and the observing system prior to making a measurement. Thus, the present formulation has some predictive ability, albeit limited. This distinguishes it fundamentally from the interpretations as they do not have such an ability; outcome of an observation is known only after it has occurred.

## 7. Concluding Remarks

An extension of Hamilton's action principle in the framework of Weyl's gauge transformations constitutes the basic founding element of the formulation presented in this paper. Fundamental underlying principles as stated in (3) and (4) of the classical and quantum mechanics, respectively, show that they are more closely related to each other than known previously and both are intertwined with the gauge transformations.

Observed impact of the method of a physical measurement on its outcome is formulated in terms of the assigned gauges determined by both, the observing and the observed system. Thus, the assigned gauges that were largely ignored, wasted, in earlier similar and not so similar formulations are used effectively in the present article, to develop an improved understanding of quantum measurement differing radically from the views in the existing interpretations of quantum mechanics, which are interpreted here in terms of the present formulation. Here an observation pertaining to a collection of particles is understood to a large extent from the properties of the observed and the observing systems prior to making a measurement, sharply contrasting it with the original formulation of quantum mechanics and its interpretations.

Feynman's path integral formulation of quantum mechanics is adjusted in conformity with the extended action principle, which defines the particle trajectories augmenting the standard formulation of quantum mechanics. The resulting wavefunction admits a geometrical interpretation as an aggregate of the Weyl lengths acquired by a unit vector transported along a collection of trajectories providing a clearer view of the wavefunction than its earlier understandings. It is shown further, that while the wavefunction is affected by the assigned gauges, Born's probability density is a uniquely defined gauge invariant quantity but at the expense of the observable information contained in the gauge factors, assigned as well as the multiplicative essential. Thus, the wavefunction and the probability density supplemented with the trajectories defined by (5) describe the behavior of a system more completely than otherwise.

Thus, the present formulation is radically at variance with the formulation and interpretations of quantum mechanics in its understanding of the observed phenomena.

Present formulation suggests and provides foundation for further investigations as follows: Although the estimates of Sec. 3 constitute a significant improvement over similar long-standing estimates in literature, further improvement is desirable, which can be done by specific calculations. Also, the discussion and conclusions of Secs. 5 and 6 are based on an analysis of the double slit experiment considering only the classical physical trajectories. Scope of the results can be enlarged by adopting the present formulation for more general cases and including all physical trajectories.

## References

1. Weyl, H. (trans. Brose, L.S.): Space-Time-Matter. Dover, New York (1951) Chs. II, IV.
2. London, F.: Quantenmechanischen deutung der theorie von Weyl. Z. Phys. 42, 375-389 (1927)
3. Adler, R.J.: Spinors in a Weyl geometry. J. Math. Phys. 11, 1185-1191 (1970)
4. Wheeler, J.T.: Quantum measurement and geometry. Phys. Rev. D 41, 431-441 (1990)
5. Feynman, R.P., Hibbs, A.R.: Quantum Mechanics and Path Integrals. McGraw-Hill, New York (1965)
6. Schulman, L.S.: Techniques and Applications of Path Integration. Wiley, New York (1981)

7. Kleinert, H.: Path Integrals in Quantum Mechanics, Statistics, Polymer Physics, and Financial Markets. World Scientific, Singapore (2009)
8. Nelson, E.: Quantum fluctuations. Princeton University Press, Princeton (1987)
9. Vatsya, S.R.: Gauge theoretical origin of mechanics. Can. J. Phys. 67, 634-637 (1989)
10. Vatsya, S.R.: Mechanics of a particle in a gauge field. Can. J. Phys. 73, 85-95 (1995)
11. Vatsya, S.R.: Current State of Quantum Theory. Phys. Astron. Int. J. 1(4): 00021. DOI: [10.15406/paij.2017.01.00021](10.15406/paij.2017.01.00021) (2017)
12. Aharonov, Y., Bohm, D.: Significance of electromagnetic potentials in the quantum theory. Phys. Rev. 115, 485-491 (1959)
13. Wu, T.T., Yang, C.N.: Concept of non-integrable phase factors and global formulation of gauge fields. Phys. Rev. D 12, 3845-3857 (1975)
14. Atkins, P.W., Friedman, R.S.: Molecular Quantum Mechanics (third ed.). Oxford University Press, New York (1997)
15. Kim, Y.H., Yu, R., Kulik, S.P., Shih, Y., Scully, M.O.: Delayed choice quantum eraser. Phys. Rev. Lett. 84, 1-5 (2000)
16. Chambers, R.G.: Shift of an electron interference pattern by enclosed magnetic flux. Phys. Rev. Lett. 5, 3-5 (1960)
17. Tonomura, A., Osakabe, N., Matsuda, T., Kawasaki, T., Endo, J.: Evidence of Aharonov-Bohm effect with magnetic field completely shielded from electron wave. Phys. Rev. Lett. 56, 792-795 (1986)
18. Vatsya, S.R.: Formulation of spinors in terms of gauge fields. Found. Phys. 45, 142-157 (2015)
19. Feynman, R.P.: Mathematical formulation of the quantum theory of electromagnetic interaction. Phys. Rev. 80, 440-457 (1950)
20. Bohm, D.: Model of the causal interpretation of quantum theory in terms of a fluid with irregular fluctuations. Phys. Rev. 96, 208-216 (1954)
21. Everett, H.: Relative state formulation of Quantum Mechanics. Rev. Mod. Phys. 29, 454-462 (1957)